\documentclass{article}
\usepackage{graphicx} 
\usepackage{authblk} 

\usepackage{titlesec}

\usepackage[letterpaper,top=2cm,bottom=2cm,left=2.5cm,right=2.5cm,marginparwidth=1.75cm]{geometry}

\usepackage{amsmath}    
\usepackage{url}
\usepackage{graphicx}   

\usepackage{csvsimple}  

\usepackage{tabularx}
\usepackage{supertabular}

\usepackage{setspace}

\usepackage{appendix}   

\usepackage{booktabs}   
\usepackage{multicol}   

\usepackage{etoolbox}   

\usepackage{siunitx}    

\usepackage{titlesec}
\titleformat{\section}{\small\bfseries}{}{0em}{}
\titleformat{\subsection}{\small\bfseries}{}{0em}{}

\title{\textbf{Revisiting Network Value: Sublinear Knowledge Law}}

\author[1]{Xinbing Wang\thanks{Corresponding to: \texttt{xwang8@sjtu.edu.cn}}}
\author[1]{Luoyi Fu}
\author[1]{Huquan Kang}
\author[1,2]{Zhouyang Jin}
\author[1]{Lei Zhou}
\author[3]{Chenghu Zhou}
\affil[1]{Shanghai Jiao Tong University, Shanghai, China}
\affil[2]{Nanjing University of Aeronautics and Astronautics, Nanjing, China}
\affil[3]{State Key Laboratory of Resources and Environmental Information System, Institute of Geographic Sciences and Natural Resources Research, Chinese Academy of Sciences, Beijing, China}

\date{} 

\begin{document}

\maketitle

\titleformat{\section}{\centering\bfseries}{}{1em}{}

\section{Abstract}

The network illustrates the interrelationships among diverse entities and has attracted considerable attention. A wide range of applications makes the network a popular modeling tool, including but not limited to social networks\cite{tang2009social}, business administration\cite{burt2000network}, and city construction\cite{Barthelemy2017TheSA}. In these fields, quantifying network value is essential. By examining network value, companies gain insights into consumer behavior, allowing them to refine marketing strategies accordingly\cite{katona2011network}. Similarly, investigating network value can assist urban planners in better understanding the utilization and demands for public facilities, ultimately optimizing city construction\cite{batty2013big}. To better understand network value, researchers have established some influential laws that describe network value in terms of the number of neighbors, edges, and subgraphs.

\textbf{Sarnoff's Law.} 
In the 1940s, David Sarnoff proposed that the value of a broadcasting network is directly proportional to the number of nodes (audience), which is also known as Sarnoff's law\cite{SWANN2002417}. For example, in the case of TV programs, the network value increases linearly with the number of the audience, because the growing number of audiences allows advertisers to access more potential customers. This results in more advertisers, increased revenues, and higher broadcast media demand. Formally, for a network with $n$ nodes, the communication value is $\Theta(n)$. This law was initially applied to the film industry, later extended to television, and usually represents one-way communication. Broadcasting can only transmit messages unidirectionally to users, but cannot spread information within users.

\textbf{Metcalfe's Law.} 
With the conferral of the Turing Award on Robert Metcalfe in 2022, Metcalfe's law\cite{metcalfe1976ethernet} has regained attention. Metcalfe's law was proposed on the background of the increasing number of Ethernet users and growing attention to the interconnection value of networks. Metcalfe argues that if the value of each node (terminal) is equal, the value of a network is proportional to the number of edges. One of the classic illustrations of Metcalfe's law lies in communication networks, where a network with $n$ users can provide interconnection value proportional to approximately $n^2$, i.e. $\Theta(n^2)$, as each user can communicate with the other $n-1$ users in the network. Despite the many challenges associated with the development of the network, such as network scale, connection quality, and network design, Metcalfe's law remains applicable in some scenarios, especially in the field of cloud computing\cite{metcalfe2013metcalfe}. The increasing number of cloud computing users leads to the availability of more resources, which in turn attracts additional users, creating a virtuous cycle of growth. As a theory for describing the interconnection value of networks, Metcalfe's law is held in high regard and possesses significant implications for communication networks, the Internet, and social networks.

\textbf{Reed's Law.} 
In 2000, David Reed proposed Reed's law\cite{reed2001law} for group-forming networks (GFNs). He argued that network value depends not only on the number of nodes in the network but also on how the nodes are connected and organized. In a highly connected network, each node can connect to other nodes to form various subgroups. The interaction and cooperation within these subgroups can bring group value to the network. This law can effectively explain the group value of large-scale networks, particularly social networks, as the group value of the network grows exponentially with the network's scale. For a network with $n$ participants, the number of possible subgroups of participants (with at least two members in the group) is $2^n-n-1$. For a network with $n$ nodes, the group value is $\Theta(2^n)$. Reed's Law is often used to explain the value of social networks such as Facebook, as these companies leverage network effects to reach more audiences. Reed's law is not without limitations, as it may overestimate the value of a network when the number of clusters exceeds human cognitive limits in larger networks.

\begin{figure}
    \centering
    \includegraphics[width=0.5\textwidth]{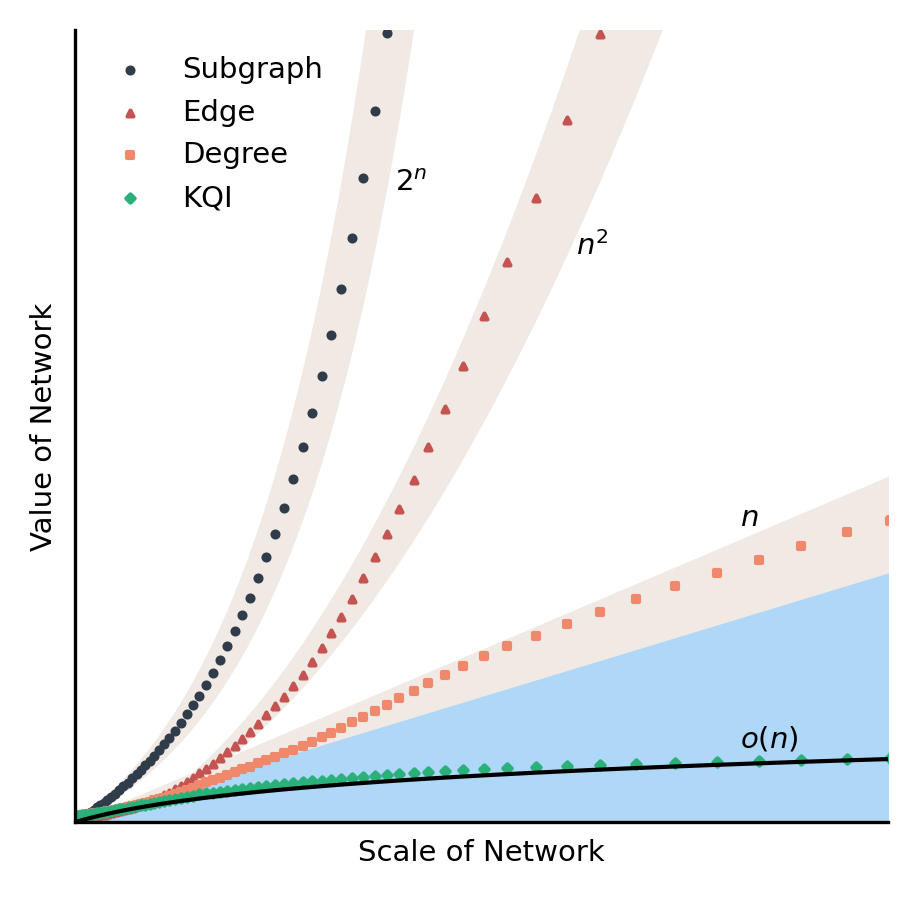}
    \caption{Sarnoff's law, Metcalfe's law, Reed's law, and sublinear knowledge law for measuring the value of different perspectives in citation networks. The figure portrays the assessment of citation networks value across different scales from diverse viewpoints, including measures of group value (represented by black circles indicating subgraph quantity), interconnectivity value (represented by red triangles signifying edge quantity), dissemination value (represented by orange squares indicating degree), and knowledge quantity (represented by green diamonds corresponding to the network's KQI). Considering the significant differences in magnitude among the degrees, edges, subgraphs, and KQI within a network, proportional scaling was implemented to achieve better visualization while maintaining accuracy in growth trends.}
    \label{fig:example}
\end{figure}

\vspace{1em}

The three laws were first proposed separately to represent network value in various contexts. In this study, we utilize the citation network constructed from the Deep-time Digital Earth (DDE)\cite{acemap} academic literature to identify the coexistence of three laws in citation networks for the first time. In a citation network, each node represents a paper, and each directed edge represents a citation relationship. The edges originating from a particular node in the citation network represent the direct influence of that paper on other papers, which indicates a linear relationship between knowledge dissemination and network scale (Figure 1). And the transmission of knowledge between papers is demonstrated by the propagation value of the citation network. Metcalfe's law is reflected in citation networks by the number of edges, which represents the total number of citations for all papers and is proportional to the square of the number of papers (Figure 1). As a result of the rapid rise in citation counts, many of the current citation-based evaluation criteria exhibit inflationary expansion over time. Reed's law can be interpreted as the number of subgraphs in a graph (Figure 1), which represents the possible literature set that could be formed by combining papers in DDE, reflecting the diversity and possibility of creativity. A literature set can construct a comprehensive knowledge system through certain combinations, which manifests the group value of citation networks.

Furthermore, we have discovered a novel law in academic citation networks, namely the sublinear knowledge law, utilizing the Knowledge Quantification Index (KQI)\cite{wang2023quantifying}, which fills a gap in network value laws.

\textbf{Sublinear Knowledge Law.} 
The previous three laws cannot reflect the amount of knowledge contained within the network. Networks are capable of storing and transmitting vast amounts of knowledge as a structure. This concept can be applied to neural networks, which are also a type of network structure. For example, the neural network-based natural language processing model ChatGPT has recently garnered considerable attention due to its extensive knowledge capacity. Therefore, it is imperative to quantify the knowledge contained within a network, which we refer to as knowledge value.

Recently, the proposal of the knowledge quantification index (KQI)\cite{wang2023quantifying} has made it possible to measure the knowledge amount contained in the network. The KQI quantifies knowledge from the perspective of information structurization and uses the extent of disorder differences caused by hierarchical structure in the citation network to represent knowledge in the literature. Using KQI to calculate the amount of knowledge contained in each yearly network snapshot of the citation network constructed based on the DDE academic literature. We found that the knowledge contained in the network grew sublinearly as the network size increased. That is, for a network with $n$ nodes, the amount of knowledge contained in it is $o(n)$ (Figure 1). The sublinear knowledge law illustrates the disparity between the growth of the network size and the growth of the amount of knowledge contained in the network, meaning that a large amount of literature does not yield a large amount of knowledge. It also implies that the current phenomenon of low efficiency in knowledge acquisition has plenty of room for improvement. At the same time, the essence of the sublinear knowledge law description is also similar to the scaling law in neural language models\cite{kaplan2020scaling}, where the performance of the model grows sublinearly with the number of parameters of the model. As the model parameters increase, the model performance improves relatively slowly.

In conclusion, the four laws mentioned above measure network value from distinct perspectives. Sarnoff's law pertains to the propagation value of the network, with emphasis on the value of its edges. Metcalfe's law pertains to the interconnection value of the network and focuses on the value of connections between pairs of nodes. Reed's law pertains to the group value of the network and highlights the value of clusters composed of multiple nodes. Here we discover the sublinear knowledge law, which provides a novel perspective regarding network value through knowledge contained within its structure. The sublinear knowledge law indicates that knowledge growth is very slow compared to the growth rate of network size and to the growth rates outlined in the three traditional laws.

\bibliographystyle{unsrt}
\bibliography{reference}

\begin{thebibliography}{10}

\bibitem{tang2009social}
J.~Tang, J.~Sun, C.~Wang, and et~al.
\newblock Social influence analysis in large-scale networks.
\newblock In {\em Proceedings of the 15th ACM SIGKDD international conference
  on Knowledge discovery and data mining}, pages 807--816, 2009.

\bibitem{burt2000network}
Ronald~S Burt.
\newblock The network structure of social capital.
\newblock {\em Research in organizational behavior}, 22:345--423, 2000.

\bibitem{Barthelemy2017TheSA}
Geoff Boeing.
\newblock The structure and dynamics of cities: Urban data analysis and
  theoretical modeling, 2017.

\bibitem{katona2011network}
Zoltan Katona, Peter~P Zubcsek, and Miklos Sarvary.
\newblock Network effects and personal influences: The diffusion of an online
  social network.
\newblock {\em Journal of marketing research}, 48(3):425--443, 2011.

\bibitem{batty2013big}
Michael Batty.
\newblock Big data, smart cities and city planning.
\newblock {\em Dialogues in human geography}, 3(3):274--279, 2013.

\bibitem{SWANN2002417}
GM~Peter Swann.
\newblock The functional form of network effects.
\newblock {\em Information economics and policy}, 14(3):417--429, 2002.

\bibitem{metcalfe1976ethernet}
Robert~M Metcalfe and David~R Boggs.
\newblock Ethernet: Distributed packet switching for local computer networks.
\newblock {\em Communications of the ACM}, 19(7):395--404, 1976.

\bibitem{metcalfe2013metcalfe}
Bob Metcalfe.
\newblock Metcalfe's law after 40 years of ethernet.
\newblock {\em Computer}, 46(12):26--31, 2013.

\bibitem{reed2001law}
David~P Reed.
\newblock The law of the pack.
\newblock {\em Harvard business review}, 79(2):23--24, 2001.

\bibitem{acemap}
{DDE Scholar}.
\newblock \url{https://ddescholar.acemap.info/}.
\newblock Accessed: April 25, 2023.

\bibitem{wang2023quantifying}
Xinbing Wang, Huquan Kang, Luoyi Fu, Ling Yao, Jiaxin Ding, Jianghao Wang,
  Xiaoying Gan, Chenghu Zhou, and John~E. Hopcroft.
\newblock Quantifying knowledge from the perspective of information
  structurization.
\newblock {\em PLOS ONE}, 18(1):1--16, 01 2023.

\bibitem{kaplan2020scaling}
Jared Kaplan, Sam McCandlish, Tom Henighan, Tom~B Brown, Benjamin Chess, Rewon
  Child, Scott Gray, Alec Radford, Jeffrey Wu, and Dario Amodei.
\newblock Scaling laws for neural language models.
\newblock {\em arXiv preprint arXiv:2001.08361}, 2020.

\end{thebibliography}

\appendix
\newpage

\begin{table}[htbp]
    \section{Appendix}
    \caption{Statistics of paper, degree, subgraph, and KQI in yearly snapshots of DDE citation network.}\label{tab:DDE}

    \begin{multicols}{2}
        \centering
        \begin{supertabular}{ccccc}
            \hline
            Year & Paper & Degree & Subgraph & KQI \\
            \hline
            \begin{filecontents*}{table1.csv}
year,paper,degree,subgraph,KQI
1920,152782,0.257157257,$9\times 10^{45991}$,0.535084172
1921,157644,0.267444368,$4\times 10^{47455}$,0.557217896
1922,162839,0.271974159,$3\times 10^{49019}$,0.568922142
1923,167778,0.278791021,$2\times 10^{50506}$,0.584959638
1924,173048,0.288792705,$4\times 10^{52092}$,0.607043223
1925,178494,0.299253756,$1\times 10^{53732}$,0.624713438
1926,183666,0.43718489,$9\times 10^{55288}$,0.858304384
1927,189127,0.610108551,$8\times 10^{56932}$,1.099898953
1928,194359,0.631506645,$8\times 10^{58507}$,1.133444606
1929,200017,0.843783278,$1\times 10^{60211}$,1.393372629
1930,205431,0.843280712,$8\times 10^{61840}$,1.404839326
1931,210803,0.950380213,$1\times 10^{63458}$,1.547316652
1932,217038,1.099885734,$9\times 10^{65334}$,1.722131243
1933,223417,1.109763357,$2\times 10^{67255}$,1.752158059
1934,229919,1.12746663,$3\times 10^{69212}$,1.79655122
1935,236499,1.153920313,$2\times 10^{71193}$,1.83876028
1936,242975,1.188546147,$6\times 10^{73142}$,1.8904614
1937,249559,1.207101327,$6\times 10^{75124}$,1.938026581
1938,256241,1.236523429,$2\times 10^{77136}$,1.992927492
1939,262737,1.253279896,$5\times 10^{79091}$,2.037396175
1940,268631,1.269112649,$1\times 10^{80866}$,2.078168129
1941,273954,1.328686568,$2\times 10^{82468}$,2.139442724
1942,279088,1.344002608,$7\times 10^{84013}$,2.172791491
1943,283828,1.389799456,$6\times 10^{85440}$,2.245456478
1944,288608,1.414344024,$5\times 10^{86879}$,2.285478855
1945,293264,1.455882754,$2\times 10^{88281}$,2.359909901
1946,298545,1.473697433,$1\times 10^{89871}$,2.399209135
1947,303445,1.50075961,$1\times 10^{91346}$,2.454036623
1948,309130,1.529062207,$3\times 10^{93057}$,2.510249352
1949,315328,1.555110869,$2\times 10^{94923}$,2.569175899
1950,321468,1.594587953,$3\times 10^{96771}$,2.648013829
1951,327968,1.633814884,$2\times 10^{98728}$,2.730545157
1952,334553,1.67843959,$3\times 10^{100710}$,2.820862868
1953,341495,1.717843599,$2\times 10^{102800}$,2.905005057
1954,348933,1.768545824,$2\times 10^{105039}$,3.014223004
1955,355850,1.824591822,$3\times 10^{107121}$,3.120596384
1956,363521,1.880474581,$5\times 10^{109430}$,3.230359476
1957,371495,1.960742406,$1\times 10^{111831}$,3.380590731
1958,380592,2.049073549,$4\times 10^{114569}$,3.537237814
1959,390298,2.143900302,$3\times 10^{117491}$,3.714461739
1960,400772,2.242551875,$2\times 10^{120644}$,3.896967564
1961,411664,2.351672723,$2\times 10^{123923}$,4.105526817
1962,423044,2.469705279,$9\times 10^{127348}$,4.328631511
1963,435458,2.603144276,$8\times 10^{131085}$,4.560013364
1964,448669,2.773420941,$7\times 10^{135062}$,4.849931766
1965,462798,2.945630707,$1\times 10^{139316}$,5.150773303
1966,477473,3.148188484,$5\times 10^{143733}$,5.471491992
1967,492846,3.373390471,$3\times 10^{148361}$,5.8359271
1968,508108,3.586717784,$6\times 10^{152955}$,6.165783489
1969,523813,3.83101794,$3\times 10^{157683}$,6.53815347
1970,541043,4.111399648,$1\times 10^{162870}$,6.973077682
\end{filecontents*}
\csvreader{table1.csv}{}{%
            \csvcoli & \csvcolii & \num[round-mode=places,round-precision=3]{\csvcoliii} & \csvcoliv & \num[round-mode=places,round-precision=3]{\csvcolv} \\} \\
            \hline
        \end{supertabular}
        
        \begin{supertabular}{ccccc}
            \hline
            Year & Paper & Degree & Subgraph & KQI \\
            \hline
            \begin{filecontents*}{table2.csv}
year,paper,degree,subgraph,KQI
1971,556328,4.349646252,$3\times 10^{167471}$,7.30974617
1972,571664,4.584007389,$1\times 10^{172088}$,7.647541359
1973,587571,4.844651965,$3\times 10^{176876}$,8.011246633
1974,603983,5.127912541,$1\times 10^{181817}$,8.383122121
1975,621084,5.514136574,$8\times 10^{186964}$,8.850324959
1976,638249,5.850053819,$1\times 10^{192132}$,9.253184858
1977,656394,6.230561218,$2\times 10^{197594}$,9.683121411
1978,675438,6.643404724,$1\times 10^{203327}$,10.12792201
1979,695615,7.214382956,$1\times 10^{209401}$,10.69722655
1980,716445,7.711913685,$3\times 10^{215671}$,11.17265646
1981,738937,8.256017766,$2\times 10^{222442}$,11.64871864
1982,761665,8.771738231,$1\times 10^{229284}$,12.0983069
1983,784602,9.45376127,$5\times 10^{236188}$,12.64076625
1984,808918,10.06405346,$4\times 10^{243508}$,13.12417682
1985,834086,10.62428455,$8\times 10^{251084}$,13.5204648
1986,860127,11.25386832,$1\times 10^{258924}$,13.96342266
1987,886974,11.95193207,$6\times 10^{267005}$,14.41658403
1988,913995,12.62720474,$8\times 10^{275139}$,14.84027041
1989,942035,13.33315747,$6\times 10^{283580}$,15.24322498
1990,970066,14.0667769,$9\times 10^{292018}$,15.65283659
1991,999605,14.93864276,$1\times 10^{300911}$,16.12105223
1992,1029509,15.72255221,$1\times 10^{309913}$,16.51303004
1993,1060900,16.55182487,$5\times 10^{319362}$,16.91749342
1994,1093183,17.39881886,$7\times 10^{329080}$,17.30662836
1995,1126668,18.36070608,$7\times 10^{339160}$,17.74962309
1996,1161641,19.2573549,$6\times 10^{349688}$,18.13703821
1997,1195940,20.31248558,$7\times 10^{360013}$,18.54373816
1998,1230995,21.42441683,$3\times 10^{370566}$,18.95194246
1999,1266709,22.56648923,$3\times 10^{381317}$,19.35414917
2000,1303786,23.89254832,$5\times 10^{392478}$,19.77932981
2001,1341797,25.24155368,$1\times 10^{403921}$,20.17806827
2002,1382738,26.69990627,$4\times 10^{416245}$,20.60295627
2003,1424870,28.20337434,$4\times 10^{428928}$,21.0188926
2004,1469500,29.71449405,$4\times 10^{442363}$,21.42486272
2005,1514667,31.30051028,$2\times 10^{455960}$,21.82659532
2006,1567014,33.08338917,$2\times 10^{471718}$,22.26098368
2007,1621004,35.06709175,$7\times 10^{487970}$,22.71260691
2008,1681158,37.20555534,$1\times 10^{506079}$,23.17866186
2009,1746961,39.49868772,$5\times 10^{525887}$,23.65555342
2010,1824318,42.01179783,$3\times 10^{549174}$,24.14391748
2011,1908775,44.39709919,$3\times 10^{574598}$,24.59700472
2012,2012905,46.91426173,$6\times 10^{605944}$,25.06672543
2013,2126676,49.97649854,$2\times 10^{640193}$,25.61198088
2014,2248549,53.08871099,$5\times 10^{676880}$,26.11971856
2015,2377431,56.24166127,$1\times 10^{715678}$,26.60003899
2016,2522132,59.50840123,$2\times 10^{759237}$,27.07344831
2017,2673591,62.84758701,$1\times 10^{804831}$,27.5340159
2018,2826169,66.23779965,$4\times 10^{850761}$,27.96751812
2019,2981690,69.68084107,$1\times 10^{897578}$,28.38770525
2020,3148739,73.27918414,$8\times 10^{947864}$,28.81422953
2021,3305200,76.03324095,$2\times 10^{994964}$,29.17055887
\end{filecontents*}
\csvreader{table2.csv}{}{%
            \csvcoli & \csvcolii & \num[round-mode=places,round-precision=3]{\csvcoliii} & \csvcoliv & \num[round-mode=places,round-precision=3]{\csvcolv} \\} \\
            \hline
        \end{supertabular}

    \end{multicols}
    
\end{table}

\end{document}